\documentclass[12pt]{JHEP}
\usepackage{amsfonts}
\def\a{\alpha}
\def\b{\beta}
\def\g{\gamma}
\def\d{\delta}

\newcommand\Appendix[1]{\def\thesection{Appendix \Alph{section}}
 \section{\label{#1}}\def\thesection{\Alph{section}}}

\def\NPB#1#2#3{{\it Nucl.~Phys.} {\bf{B#1}} (#2) #3}
\def\PLB#1#2#3{{\it Phys.~Lett.} {\bf{B#1}} (#2) #3}
\def\PRD#1#2#3{{\it Phys.~Rev.} {\bf{D#1}} (#2) #3}

\def\JHEP#1#2#3{{\it J. High Energy Phys.} {\bf#1} (#2) #3}
\newskip\humongous \humongous=0pt plus 1000pt minus 1000pt
\def\caja{\mathsurround=0pt}
\def\eqalign#1{\,\vcenter{\openup1\jot \caja
        \ialign{\strut \hfil$\displaystyle{##}$&$
        \displaystyle{{}##}$\hfil\crcr#1\crcr}}\,}
\newif\ifdtup

\newcommand{\be}{\begin{equation}}
\newcommand{\ee}{\end{equation}}
\newcommand{\ba}{\begin{eqnarray}}
\newcommand{\ea}{\end{eqnarray}}
\newcommand{\ban}{\begin{eqnarray*}}
\newcommand{\ean}{\end{eqnarray*}}

\title{Effective Action of Matter Fields
in \\ Four-Dimensional String Orientifolds} 
\author{P. Bain \\ 
Laboratoire de Physique Th{\'e}orique de l'Ecole Normale
Sup\'erieure\footnote{Unit\'e mixte du CNRS et de l'ENS, UMR 8549}\\ 
24 rue Lhomond, F-75231 Paris,  France\\
{\tt bain@physique.ens.fr}}
\author{M. Berg\footnote{on leave of absence from  
Center for Relativity, Department of Physics,
University of Texas at Austin, 78712 Austin,
TX, USA} \\ 
D{\'e}partement de Math{\'e}matiques et
Applications de l'Ecole Normale Sup\'erieure\footnote{Unit\'e mixte
du CNRS et de l'ENS, UMR 8553}\\
45 rue d'Ulm, F-75230 Paris, France\\
{\tt berg@dma.ens.fr}}
\abstract{We study various aspects of the K\"ahler metric for matter 
fields in ${\cal N}=1, 2$ orientifold compactifications of type IIB 
string theory. 
The result has
an infrared-divergent part which reproduces the field-theoretical
anomalous dimensions,
and a moduli-dependent part which comes from ${\cal N}=2$ sectors
of the orientifold. 
For the ${\cal N}=2$ orientifolds, we also compute the disk amplitude for
two matter fields on the boundary and a twisted 
closed string modulus in the bulk. 
Our results are in agreement with supersymmetry: the 
singlet under the $SU(2)_R$ $R$-symmetry has vanishing coupling,
while the coupling of the $SU(2)_R$ triplet does not vanish.}
\preprint{LPTENS-00-12 \\ DMA-00-12 \\ hep-th/0003185}

\begin{document}
\section{Introduction}

In the past few years, due to the improved
understanding of the role of D-branes 
in string theory, four dimensional ${\cal N}=1$ Type IIB orientifold 
compactifications \cite{ABPSS,AFIV} have received renewed attention.
Compared to their weakly-coupled heterotic counterparts,
which have been more thoroughly explored \cite{Ka,DKL},
these models offer added flexibility, since the tree-level relations 
between gauge and string couplings or compactification and string 
scales are non-universal. In particular, these models play an
important part in brane-world scenarios (see \cite{BA} and references
therein). 

The purpose of this paper is to continue the work of \cite{BF,ABD}
on the determination of some parts of the effective action for these 
orientifold compactifications.  
While \cite{ABD} discussed the 
gauge couplings, this article will focus on the study of the couplings
of the matter fields.
Let us first 
recall some general facts about the effective action of a
four-dimensional field theory with ${\cal N}=1$ or ${\cal N}=2$
supersymmetry, emphasizing the main characteristics of interest here. 
More detail can be found in \cite{E}, for instance.

The bosonic part of the effective Lagrangian with at most two
derivatives is given by the following expression: 
\ban{\eqalign{
{\cal L}_{\rm eff} = -\frac{R}{2\kappa^2} + \frac{1}{2 g_a^2(z)}{\rm
tr}_a(F_{\mu\nu}F^{\mu\nu}) &+ \frac{\Theta_a(z)}{16\pi^2}{\rm
tr}_a(F_{\mu\nu}\tilde{F}^{\mu\nu}) \cr
&+ \frac{1}{2} G_{ij}(z)D_\mu z^i D^\mu z^j + V(z)
}}
\ean
where $z$ are the scalar fields which parametrize a K\"ahler manifold
of metric $G_{ij}$, and $V(z)$ is their potential. 
 For ${\cal N}=1$ supergravity, this effective action is
completely defined by the following functions~:
\vspace{-2mm}
\begin{itemize}
\item the K\"ahler potential $K(z,\bar{z})$ which determines the
scalar metric $G_{i\bar{\jmath}}= \partial_i \partial_{\bar{\jmath}}
K$. For ${\cal N}=1$ type I orientifolds without D5-branes, the matter
field dependent part of the 
tree-level K\"ahler potential reads \cite{AFIV} 
$K = - \sum_{i=1}^3 {\rm log}({\rm Im} \, T^i +
|\phi^i|^2/2)$ where $T^i$ is the K\"ahler structure and $\phi^i$ is
the scalar matter field associated to the  $i^{\rm th}$ torus, 
\vspace{-2mm}
\item the analytic superpotential $W(z)$ which determines the part of
the scalar potential associated to the F-terms~:
$V_F=G^{i\bar{\jmath}} \partial_i W \partial_{\bar{\jmath}} W$,
and which is not renormalized in perturbation theory,
\vspace{-2mm}
\item the analytic function $f_a(z)$ which gives the gauge couplings
and the theta angles as $f_a(z) = \Theta_a(z)/8\pi^2+i/g^2_a(z)$ ,
\vspace{-2mm}
\item the Fayet-Iliopoulos (FI) D-terms, due to the presence of anomalous
$U(1)$ factors of the gauge group. In type I compactifications, the
anomaly-cancellation mechanism involves twisted Ramond-Ramond (R-R)
fields and gives rise to D-term contributions to the
scalar potential at tree-level.
\end{itemize}
\vspace{-2mm}
The calculation of the 
analytic function $f_a$ was the subject of ref.\ \cite{ABD},
where the tree-level couplings to twisted moduli and one-loop
renormalization were extracted from annulus and M\"obius strip
diagrams, evaluated in a background magnetic field. Here, we will
extend these results to other parts of the effective action, and in
particular, we will find the one-loop renormalization of the K\"ahler
metric of the matter fields charged under the gauge groups. However,
we will perform direct calculations of the relevant scattering
amplitudes, rather than using the background field method, for a
reason we will explain below. The new results are as follows:
\vspace{-2mm}
\begin{itemize}
\item in ${\cal N}=1$ sectors, the string oscillator modes do not
decouple, and cut off the one-loop amplitude at the string scale
$M_{\rm S}$.
The infrared (IR) behavior of the
one-loop string amplitude is governed by the field theoretical
anomalous dimensions (example of the $T^6/{\mathbb Z}_3$ model),
\vspace{-2mm}
\item in ${\cal N}=2$ sectors, only BPS states contribute to the D9-D9
annulus and to the M\"obius strip amplitudes, and give
rise to moduli-dependent
threshold corrections for the matter field associated
with the untwisted direction (example of the $T^6/{\mathbb Z}_6^\prime$
model),
\vspace{-2mm}
\item  one-loop induced FI terms are absent for
models with D5-branes and ${\cal N}=2$ sectors, generalizing the
result of \cite{P}.
\item tree-level D-terms, given by the coupling of twisted closed
string states to bilinears in matter fields, contribute 
to the scalar potential. We have calculated them only in the context
of ${\cal N}=2$ compactifications, where the twisted moduli space is
simpler.
\end{itemize}
Compared to ${\cal N}=1$ compactifications, ${\cal N}=2$ supersymmetry
imposes further restrictions on the effective action; in particular,
at two-derivative order, there is no mixing between the 
hypermultiplets and the vector multiplets \cite{VP}\footnote{Except
those dictated by gauge symmetry \cite{E}}. Moreover, for
${\cal N}=2$ type I compactifications, the four-dimensional dilaton
belongs partly to a vector multiplet and partly to a hypermultiplet
\cite{ABFPT}. This should be contrasted with what happens in heterotic
string theory, where the dilaton is in a vector multiplet, or with
type II compactifications, where it is in a hypermultiplet. 

The paper is organized as follows; in section two, we describe the 
methods used to derive the tree-level couplings and one-loop corrections 
of the matter field metric for a general orbifold compactification. 
In particular, we give general expressions for the tree-level 
amplitude involving one closed, twisted NS-NS field and two 
open-string matter fields and needed to extract the FI D-terms, as outlined
in the appendix of \cite{DM}. We also present the one-loop, two-point
functions needed to obtained the one-loop corrections.
Then, in section three, we apply these methods to 
$K3\times T^2$ orientifolds, showing which twisted moduli are
effectively involved in the FI couplings. In section four, we discuss
the one-loop renormalization of the K\"ahler metric of matter fields
in ${\cal N}=1$ compactifications;
we verify that, for the ${\cal N}=1$ ${\mathbb Z}_{3}$ model, string theory
reproduces the field theoretical anomalous dimensions. Finally, we
study the ${\cal N}=1$ ${\mathbb Z}_{6}^\prime$ model 
and we comment on the effective field theory interpretation.

\section{General methods}
\label{sec:general}

We will consider four-dimensional ${\mathbb Z}_{N}$ orientifolds obtained by 
orbifolding the six-torus $T^6$ by the twist operator $\theta = 
e^{2\pi i v_{i} J_{i}}$, with $J_{i}$ the generator of the rotation in 
the $i$-th complex plane. For a ${\mathbb Z}_N$ orientifold, $\theta^N=1$. 
Here $v_i=(v_1, v_2, v_3)$ is known as the twist vector.
The twist $\theta^k$ also acts on the $n 
\times n$ Chan-Paton  factors: 
this action is realized by $n \times n$ matrices, $\gamma^k$. 
We call the three complex coordinates of the six-torus
$Z^i \equiv (X^{2i+2} + i X^{2i+3})/2$ for  $i=1,2,3$, 
and denote by $\phi^i$ the open string massless states associated to
these directions,
which correspond to Wilson lines for the D9-branes, or describe
transverse positions of the D5-branes.
The action of the orbifold on their Chan-Paton wave 
functions $\lambda_{i}$ is $\gamma^k \lambda_{i} (\gamma^k)^{-1}= 
e^{-2\pi i v_{i}} \lambda_{i}$ in order to obtain an invariant state.
We will use latin indices $i,j,\cdots$ for compact complex
coordinates and greek letters $\mu,\nu\cdots$ for four-dimensional
spacetime coordinates.   

{
Denoting by $G_i$ the metric of the $i^{\rm th}$ torus $T^i$ and by
$V_i=\sqrt{G_i}M_{\rm S}^2$ its volume, the 
gauge coupling constant on D9-branes 
is given by the inverse of the imaginary part
of $S = a^{\rm R-R}+i V_1 V_2 V_3\, e^{-\Phi_{10}}$. We also recall that 
the four-dimensional and ten-dimensional dilatons are related by
$e^{-2\Phi_4}=e^{-2\Phi_{10}}V_1 V_2 V_3 M_{\rm S}^{-6}$.
}

\subsection{Background field method versus ``dynamical branes''}

In \cite{ABD}, the tree-level couplings to the closed twisted fields and 
the one-loop renormalization of the gauge couplings were extracted 
from a one-loop vacuum energy calculation in a background 
magnetic field, for various orientifold compactifications of type IIB 
string theory. The effect of this background field is to modify the 
boundary conditions for the open string \cite{ACNY}. 
Unfortunately, this method is not so
useful for {\it twisted} complex coordinates, as we will explain below.
First, we show
that for coordinates left untwisted in some specific 
sectors, one can use a variant of the background field method, as follows. 
If $X^4$ and $X^5$ are the coordinates left untwisted, one
takes the T-dual along one of these directions, say for instance $X^4$. 
This duality transforms D9-branes into D8-branes. Then, one gives 
an angle $\theta$ to one of these D8-branes in the $X^1 X^4$ plane.
The boundary conditions for a string stretched between this 
tilted D-brane and an untilted one become
\ban
\left\{
\begin{array}{l}
\partial_\sigma X^1(0,\tau) = 0 \\
X^4(0,\tau) = 0 
\end{array} \right. \qquad
\left\{ \begin{array}{l}
\partial_\sigma X^1(\pi,\tau) + 
\partial_\sigma X^4(\pi,\tau) \tan \theta  = 0 \\
X^1(\pi,\tau) \tan \theta - X^4(\pi,\tau) = 0 \; .
\end{array} \right.
\ean
If we calculate the partition function in this background, the result
turns out to be the same as the one obtained in \cite{ABD} for the
gauge fields.  

Now consider instead T-dualizing in four compact, twisted directions to 
turn the D9-branes into D5-branes.
For a  twisted coordinate, giving an expectation value
(linear in an untwisted coordinate) to the associated field means
pulling the brane away from a fixed point, or in field theory language,
moving on 
the Higgs branch of moduli space. We can pull branes
away from the fixed point of a ${\mathbb Z}_N$ orbifold 
only in certain combinations of
$N$ branes into a ``dynamical
brane'', which has no total charge
under the twisted sector of the orbifold. To be precise, a dynamical brane
away from a fixed point is made out of $N$ copies of the brane
under the action of the orbifold group, 
${\mathbb Z}_N$. (For the orientifold, we also have to include their images under
world-sheet parity $\Omega$ \cite{GP,GJ}.) Labelling these branes by
Chan-Paton indices
$i=1,\cdots,N$, their positions are given by $X(i)$ and the orbifold
group acts as $\theta(X(i))=X(\gamma(i))$. Therefore, the Chan-Paton
representation of $\theta$ on these branes is the permutation matrix~:
\ban
\label{cpbulk}
\gamma~=~ 
\left( 
\begin{array}{ccccc}
\; 0 \; & \; 1 \; & \; 0 \; & \cdots & \; 0\;  \\
0 & 0 & 1 & \cdots & 0 \\
  &   &   & \vdots & \\
1 & 0 & 0 & \cdots & 0 
\end{array} 
\right)
\ean
We see that the dynamical brane is in the regular representation
$R$ of the orbifold group. Now, it is easy to show that the boundary state
associated to the representation $R$ is uncharged under the
closed string twisted sector. Such boundary states have been
constructed in \cite{T}. Since ${\rm tr}_R(\gamma^k)=0$ for
$k=1,\cdots,N-1$, the contributions of the twisted sectors to the
boundary state describing a brane in the regular representation
vanish. Therefore, branes away from fixed points
have no couplings to the closed string twisted sector. In
particular, this argument can be applied to branes in the magnetic
field of \cite{ABD}: when pulled away from a fixed point,
these branes no longer have tree-level couplings between the twisted
moduli and the gauge kinetic term. This also resolves an
apparent paradox about the contribution of
the classical action to these couplings. 
One could try to argue that for branes at a
nonzero distance $|\phi|$ from an orbifold fixed point,
the coupling would be suppressed by the classical action as
${\rm exp}(-|\phi|^2/\alpha^\prime)$.
However, we
know that such a term is not compatible with four-dimensional ${\cal
N}=2$ supersymmetry\cite{VP} since it involves a two-derivative
coupling between hypermultiplets and vector multiplets.
Happily, we have seen that this coupling is in fact absent 
for branes away from the fixed point, so there is no paradox.
By the same token, we see that the background field method applied to
the twisted coordinates can only give contributions from the untwisted
sector of the orbifold. To obtain the twisted sector
contributions to the couplings and renormalizations of these matter
fields, we need to directly compute the relevant scattering amplitudes.

\subsection{One-loop two-point function}

To extract the one-loop renormalization of the wave function of the
charged matter field, we calculate the even spin-structure 
part of the annulus and M\"obius strip amplitudes 
with two open string vertices polarized in a twisted complex 
direction and inserted on the boundary, which is 
stuck on a D9-brane. The annulus with both
ends on D9-branes reads: 
\[
{\cal A}_{99}(\phi^i,{\phi^{\bar{\imath}}})=-\;\frac{1}{4N}\int_0^1 
\frac{dq}{q} \int_0^q \frac{dz}{z} \int 
\frac{d^{4}p}{{(2\pi)}^{4}} \sum_{k=0}^{N-1} 
{\rm Tr}\Bigl[\theta^k V(\xi_1,p_1;z) V(\xi_2,p_2;q) \,  q^{L_0}\Bigr]
\]
where $L_0 = (p_{\mu}p^{\mu}+m^2)/2$ since we take $\alpha'=1/2$.
We will use the RNS formalism. The relevant vertex
operators in the zero ghost picture are
\ba
V(\xi,p\,;z) = \lambda \; \xi_i (\partial X^i + i(p \cdot \psi)\psi^i)\;
e^{ip \cdot X(z)} \label{openv}
\label{vertex}
\ea 
where $\xi_i$ is the polarization vector of the scalar $\phi^i$,
and $\lambda$ is the Chan-Paton factor as above. 
Since we are interested in the one-loop renormalization of the kinetic 
terms of the matter fields, we will consider only the ${\cal O}(p^2)$ 
contribution to the two-point function. However, due to mass-shell 
conditions and momentum conservation, this amplitude vanishes. 
To extract information from this two-point function, we
should relax 
one of these two conditions. To see which one, we recall that 
vertex operators of physical states must be BRST-invariant. For
(\ref{vertex}), the conditions are $p \cdot \xi_i = p \cdot p=0$. On the
other hand, momentum conservation comes from the integration of the
zero modes which gives a function $\delta({\scriptstyle \sum}_i
p_i)$, where the $p_i$ 
are the momenta of the external legs. Since the integration of 
zero modes is independent of the BRST conditions, we
can relax momentum conservation and still have physical
string amplitudes: $\delta \equiv p_1 \cdot p_2 \neq 0$.  
Although not completely justified\footnote{This method has also been
used in 
\cite{ABFPT} to compute the renormalization of the Planck scale in
orientifold compactifications. As explained there, a correct
procedure is to start with a three-point amplitude with a U-modulus or the
dilaton and the two other fields (for them gravitons,
for us scalar fields) which are on-shell but have complex
momenta. See also \cite{KK} for another alternative and more justified
way to calculate these corrections.},
we will see in the following that this 
procedure gives results which agree with the effective field
theory description and with the heterotic counterpart \cite{AGNT}, when
available.
Moreover, for the leading, $\delta$ independent term, one can justify
the calculation by factorizing a four-point function, as in
\cite{DIS}.
Finally, our method is in essence very similar to that used in
ref.\ \cite{Ka}.

Doing the contractions, the annulus diagram reduces to:
\ba
{\cal A}_{99}(\phi^i,{\phi^{\bar{\imath}}}) &=& - \frac{1}{2N}\;
 {p_1}_\mu {p_2}_\nu \; \xi^i {\xi^{\bar{\imath}}}\; 
\sum_{k=0}^{N-1} {\rm tr}
({\g}_9^{k}{\lambda}_1^\dag {\lambda}_2)\;{\rm tr} ({\g}_9^{k}) 
\nonumber \\
&& \; \times \;  \int_0^{i \infty} d\tau 
  \sum_{\a,\b=0,1/2 \atop {\rm even}}  {1\over 2}\, \eta_{\a,\b}\
Z_{4}^{\a,\b}(\tau) 
\ Z^{\a,\b}_{{\rm int},\; k}(\tau)   \\
&& \; \times \; \int_0^{\tau} d\nu\; 
e^{-\delta \langle X(z) X(q)\rangle } \;
\langle \psi^\mu(z)\psi^\nu(q)\rangle ^{\a,\b}\; 
\langle \psi^i(z)\psi^{\bar\imath}(q)\rangle ^{\a,\b} 
\nonumber
\label{annulus99}
\ea
where we have introduced $q=e^{2\pi i \tau}$, $z=e^{2\pi i
\nu}$ and $\eta_{\alpha, \beta} = (-1)^{2\alpha + 2\beta + 4 \alpha
\beta}$. The contribution of the zero modes and oscillators of the 
spacetime coordinates and ghosts to the partition function, denoted
 $Z_{4}^{\a,\b}(\tau)$, and of the compact coordinates,
$Z^{\a,\b}_{\rm int}(\tau)$, are 
given in the appendix for all the models we will consider in this
paper. The correlation functions are also given in this appendix.
As usual, the M\"obius strip amplitude is obtained by 
shifting the modular parameter $\tau$ by $1/2$ and taking into account 
the modification of the Chan-Paton traces. 
For models with D5-branes, 
the annulus amplitude with one boundary on the D9-branes and the other 
on the D5-branes may also contribute to the renormalization of the
K\"ahler metric. 

We can formally expand the contributions of these amplitudes in powers
of the momentum as
\ban
{\cal A}(\phi^i,{\phi^{\bar{\imath}}}) = \left( \zeta_{\phi^i} + 
\tilde{\gamma}_{\phi^i} \delta + {\cal O}(\delta^2) \right) \xi^i
{\xi^{\bar{\imath}}} 
\ean
The first coefficient, $\zeta_{\phi^i}$, is the one-loop FI term while
$\tilde{\gamma}_{\phi^i}$ will give the one-loop renormalization of the
matter field metric. If there is an untwisted two-torus in a sector of
the orbifold, these coefficients can depend explicitly 
through a logarithm on its moduli,
as we will see in the following sections.

In the string frame,
reinstating the tree-level contribution and the Einstein term, the
two-derivative effective action for the matter fields reads~:
\ban
{\cal L}^{\rm (S)} = -\frac{1}{2 \kappa^2} V_1 V_2 V_3 M_{\rm S}^{-6} \, 
e^{-2\Phi_{10}}
R + 
\left( V_1 V_2 V_3 M_{\rm S}^{-6} \, e^{-\Phi_{10}} G_{i \bar{\imath}} +
\tilde{\gamma}_{\phi^i}  \delta_{i \bar{\imath}}
\right) \partial_\mu \phi^i \partial^\mu
\phi^{\bar{\imath}}   
\ean
where $G_{i \bar{\imath}}$ is the tree-level metric, which, for
correctly normalized string vertex operators, begins with $\delta_{i
\bar{\imath}}$ (see appendix B of \cite{KL} for instance).
To compare this string theory result with the field theory predictions,
one has first to go to the Einstein frame. The correct redefinition of
the metric in four dimensions is: 
\ban
\label{rescaling}
G_{\mu\nu}^{\rm (S)} = e^{2\Phi_4} G_{\mu\nu}^{\rm (E)}
\ean
After this redefinition, the Lagrangian density reads~: 
\ba
\label{oneloop}
{\cal L}^{\rm (E)} = -\frac{1}{2 \kappa^2} R + 
\left( G_{i \bar{\imath}} + \frac{\tilde{\gamma}_{\phi^i}}{{\rm
Im}\,S} \, \delta_{i \bar{\imath}}
 \right) e^{\Phi_{10}} \, \partial_\mu \phi^i \partial^\mu
\phi^{\bar{\imath}}.
\ea
Here we ignore the one-loop
universal correction to the Einstein term \cite{ABFPT}.

\subsection{Disk amplitude and tree-level couplings}
\label{sec:disk}

The tree-level couplings to closed twisted NS-NS fields can be obtained 
from a disk amplitude with two open vertices and one closed vertex. 

The amplitude involves two charged matter vertices inserted on 
the boundary of the disk and one closed twisted vertex $V(\rho, k; z)$ in the 
interior:
\ban
{\cal A} = \int \frac{d^2z \, dx_{1} \, dx_{2}}{V_{\rm CKG}}\;
{\rm Tr}\langle \, V(\xi_{1}, p_{1}; x_{1}) \; V(\xi_{2}, p_{2}; x_{2})\; 
V(\rho, k; z) \, \rangle
\ean
where $V_{\rm CKG}$ is the volume of the $PSL(2,{\mathbb R})$ 
conformal Killing group 
of the disk. This $PSL(2,{\mathbb R})$ 
invariance can be used to fix the positions 
of $V(\rho, k; z)$ and of one of the boundary operators; by a conformal 
transformation, we map the disk to the upper-half plane and choose 
$z=i$, $x_1=x$ and $x_2=-x$.  Since the total superconformal ghost 
number of the disk is $-2$, we may choose the $(-1, -1)$ picture for the 
closed string vertex:
\ba
V(\rho, k; z) = \rho_{mn} e^{-\phi(z, \bar z)} \psi^m(z) \sigma_k(z)
\tilde\psi^n(\bar z) \sigma^\dag_k(\bar{z}) e^{i k\cdot X(z, \bar z)}
\label{closedv}, 
\ea
where the $\sigma_k$, $\sigma^\dag_k$ are the ${\mathbb Z}_N$ twist fields
\cite{DFMS}. 
The relevant open string vertices are given by eq.\ (\ref{openv}).
The correlators we need to evaluate the disk amplitude are given in the
appendix. Then, it is easy to see that the amplitudes can expressed
in term of the following integral:
\ban
I(\delta, \alpha)= 2^{\delta} \int_0^\infty dx \; x^{\delta-1}
(x-i)^{-\delta+\alpha}(x+i)^{-\delta-\alpha} 
\ean
for general $\delta$ and $\alpha$,
which can be evaluated explicitly using hypergeometric
functions; the result is 
\ba
\label{int}
I(\delta, \alpha)=
2^{\delta} e^{\frac{i \pi \delta}{2}}B(\delta,\delta) \
_2F_1(\delta+\alpha,\delta;2\delta;2) 
=\sqrt{\pi} e^{-\frac{i\pi \alpha}{2}} 
\frac{\Gamma(\frac{\delta}{2})\Gamma(\frac{\delta+1}{2})}
{\Gamma(\frac{\delta+1+\alpha}{2})\Gamma(\frac{\delta+1-\alpha}{2})}
\; .
\ea
We will use this result in the following section.

\section{${\cal N}=2$ supersymmetry:  $K3\times T^2$ orientifolds}

\subsection{(No) one-loop renormalization of the hyperk\"ahler metric} 

We start with the simplest models obtained by compactifying the 
six-dimensional ${\mathbb Z}_2$ orientifold \cite{BS,GP} (and its
${\mathbb Z}_{N}$  
generalizations \cite{GJ,DP}) to four dimensions. 
For these models, the twist vector is ${\mathbf v}=(1/N,-1/N,0)$. 
Tadpole cancellation requires that we introduce 32 D9-branes and, for $N$ 
even, 32 D5-branes.
The six-dimensional ${\cal N}=(1,0)$ chiral hypermultiplet 
becomes a four-dimensional ${\cal N}=2$ hypermultiplet, while the 
vector multiplet gives an ${\cal N}=2$ vector multiplet whose complex 
scalar component comes from the directions along on the untwisted
two-torus. 
Consequently, the renormalization of the metric of these latter
scalars is related to the renormalization of the coupling
constants. One can also see this result directly from a string  
calculation, using the background field method described in the previous 
section, which is valid for untwisted coordinates.

On the other hand, the four scalar fields which belong to the 
hypermultiplet correspond to twisted coordinates and require the 
direct methods outlined above. Let us begin with the one-loop 
amplitude.  Using the correlators, the partition functions and the
theta function identity given in the appendix, one sees that the annulus
amplitude vanishes~: 
\ban
\eqalign{
{\cal A}_{99}(\phi^i,{\phi^{\bar{\imath}}}) & \!=\! \;
- \frac{\delta \; \xi^i {\xi^{\bar{\imath}}}}{8\pi^2 N} \; 
\sum_{k=0}^{N-1} 
{\rm tr}({\g}_9^{k}{\lambda}_1^\dag {\lambda}_2)\;
{\rm tr}({\g}_9^{k})\;  
(2 \sin{\pi k \over N})^2 \cr
& \int_0^{i \infty} \;
\frac{d\tau}{2\tau^2}\;\Gamma^{(2)}(\tau)
\frac{\vartheta_1(0|\tau)\;\eta(\tau)^{3\delta}}
     {\vartheta[{1/2\atop{1/2 + kv_i}}](0|\tau)}  
\int_{0}^{\tau} d\nu\; e^{i\pi \delta \frac{\nu^2}{\tau}} \; 
\frac{\vartheta[{1/2 \atop {1/2 + k v_i}}](\nu|\tau)}
     {\vartheta_1(\nu|\tau)^{\delta+1}}
= 0 \; .} 
\ean
Similarly, one can easily see that the M\"obius strip amplitude and,
for $N$ even, the D9-D5 annulus also vanish. 
Moreover, the
presence of a $\vartheta_1$ function with zero first argument shows that
this result is related to the number of supersymmetries preserved by 
the compactification. 
To compare with the effective field theory prediction, we use 
equation (\ref{oneloop}). For non-vanishing $\tilde\gamma$
coefficients, it predicts  two-derivative coup\-lings between the
fields $\phi^i$ which are in hypermultiplets and $S$ which is in a
vector multiplet. Since such terms are forbidden by ${\cal N}=2$
supersymmetry, we conclude that 
field theory also predicts the absence of one-loop renormalization of
the hyperk\"ahler metric.

\subsection{Twisted tree-level couplings and Fayet-Iliopoulos terms}

Now, we will derive the tree-level couplings of the closed string
twisted fields to the charged hypermultiplets. To do this, we first
need to recall the
structure of the twisted moduli in $T^4/{\mathbb Z}_N \times T^2$ orientifold
compactifications \cite{GJ,DP,DM}. These twisted moduli can be
interpreted as the Kaluza-Klein reduction of the ten-dimensional
fields of type IIB string theory on supersymmetric two-cycles of $K3$,
projected by the product of the orientation reversal operator $\Omega$
and an operator $J$ acting on the twisted sectors.
As explained in \cite{POL}, this operator $J$ exchanges sectors $k$
and $N-k$. For the untwisted sector, $J$ is just the identity. In type
IIB, the dilaton, the metric and the R-R 2-form are $\Omega$-even
while the NS-NS tensor, $B^{(2)}$, and the R-R scalar and 4-form are
$\Omega$-odd. Therefore, the bosonic four-dimensional states are
obtained by contracting the $\Omega$-even (odd) ten-dimensional
fields with $J$-even (odd) combinations of harmonic forms from the $k$
and $N-k$ twisted sectors.

In the twisted NS-NS sector, the massless fields are given by the
tensor product of left and right moving modes:
\ban{\eqalign{
\label{nsnstw}
&\left(
\begin{array}{c}
\psi^{\bar{1}}_{-1/2+k/N} \\
\psi^{2}_{-1/2+k/N} 
\end{array} 
\right)
\otimes
\left( 
\begin{array}{c}
\tilde{\psi}^{1}_{-1/2+k/N} \\
\tilde{\psi}^{\bar{2}}_{-1/2+k/N}
\end{array} 
\right)
|p;k,\mbox{NS-NS}\,  \rangle \ \ {\rm for }\ \  1 \leq k < N/2 \cr
&\left( 
\begin{array}{c}
\psi^{1}_{-1/2+(N-k)/N} \\
\psi^{\bar{2}}_{-1/2+(N-k)/N} 
\end{array} 
\right)
\otimes
\left( 
\begin{array}{c}
\tilde{\psi}^{\bar{1}}_{-1/2+(N-k)/N} \\
\tilde{\psi}^{2}_{-1/2+(N-k)/N}
\end{array} 
\right)
|p;k,\mbox{NS-NS} \, \rangle \ \ {\rm for }\ \  N/2 \leq k \leq N\; .
}} 
\ean
As in \cite{DM}, we decompose the rotational symmetry $SO(4)$ which
acts on the coordinates of the four-torus as $SU(2)_L\times SU(2)_R$
and define
\ban
\Psi~=~ 
\left( 
\begin{array}{cc}
\psi^1 & -\psi^{\bar{2}}   \\
\psi^2 & \psi^{\bar{1}} \\ 
\end{array} 
\right) \; .
\ean
One can classify the four twisted fields (\ref{nsnstw}) according to their
transformations under 
the $R$-symmetry group $SU(2)_R$, and define a triplet ${\rm tr}(\Psi
\vec{\sigma} \Psi^\dag)$ and a singlet ${\rm tr}(\Psi
\Psi^\dag)$. The triplet 
state is associated to the complex structure and K\"ahler deformations
of the manifold, whereas the singlet $b^{(0)}$ comes from the Kaluza-Klein
reduction of the ten dimensional $\Omega$-odd $B^{(2)}$ field on a
vanishing supersymmetric two-cycle $\Sigma_k$ of the orbifold.  
Explicitly, 
the states are
\be
\begin{array}{|c|c|}\hline
\mbox{ state } & \mbox{given by the action of} \\[0.5mm] \hline\hline 
\raisebox{-1mm}{$b^{(0)}$}  &
\raisebox{-1mm}{$\; \psi^{\bar{1}}_{-1/2+k/N}\tilde{\psi}^{1}_{-1/2+k/N}
+ \psi^{2}_{-1/2+k/N}\tilde{\psi}^{\bar{2}}_{-1/2+k/N}$}  \\[3mm] \hline
\raisebox{-1mm}{$\rho^3_k$} &
\raisebox{-1mm}{$\; \psi^{\bar{1}}_{-1/2+k/N} \tilde{\psi}^{1}_{-1/2+k/N}
- \psi^{2}_{-1/2+k/N}\tilde{\psi}^{\bar{2}}_{-1/2+k/N}$}  \\[2mm]
\rho^+_k  &\psi^{2}_{-1/2+k/N}\tilde{\psi}^{1}_{-1/2+k/N}  \\[1mm]
\rho^-_k  &\psi^{\bar{1}}_{-1/2+k/N}\tilde{\psi}^{\bar{2}}_{-1/2+k/N}
\\[2mm] \hline
\end{array}
\ee 
on the $k$-twisted NS-NS ground state
(here we have omitted the contributions of the
$N-k$-sectors to these fields).  
We use the Pauli matrices $\sigma^{\pm}=\sigma^1 \pm
i\sigma^2$.

The R-R sector gives a six-dimensional anti-self-dual twisted 2-form and a
twisted scalar, coming from the reduction of the R-R 4 and 2-forms on
$\Sigma_k$: 
$$\ ^6C_k^{(2)} = \int_{\Sigma_k} \! ^{10}C^{(4)} \; , \qquad ^6C_k^{(0)} =
\int_{\Sigma_k} \! ^{10}C^{(2)} \; .$$ 
Reducing the anti-self-dual
antisymmetric field on the two-torus gives a four-dimensional vector
and an antisymmetric tensor (or, equivalently, its scalar dual).

The four fields (the NS-NS triplet and the R-R scalar $^6C_k^{(0)}$)
which come from the Kaluza-Klein reduction of the $\Omega$-even sector
fill out a 
hypermultiplet, while the singlet and its R-R partner $^6C_k^{(2)}$
give a ${\cal N}=2$ four-dimensional vector-tensor multiplet.

Since according to \cite{VP}, there are no couplings
between hypermultiplets and vector multiplets up to second order in
derivatives, we expect that the amplitude with two charged
hypermultiplets and the twisted singlet vanishes. On the other hand,
the triplet can couple to the charged hypermultiplets and, indeed,
it will correspond to an FI term as argued in
\cite{DM}. We will now verify this claim by a direct calculation,
using the method described in section \ref{sec:disk}.

Using (\ref{int}), the disk amplitude with insertion of the singlet in
the bulk and two charged open strings vertices (polarized in the
twisted directions) on the boundary vanishes: 
\ban
\eqalign{
{\cal A}(b^{(0)}_k, \phi_1, \phi_2) = b^{(0)}_k \xi^i {\xi^{\bar{\imath}}}
{\rm tr}({\g}_9^{k}{\lambda}_1^\dag {\lambda}_2)
\Bigl(&         \delta I(\delta, 2k v_i-1)
 -i(\delta - 1 + k v_i)I(\delta-1, 2k v_i) \cr
      &       +i k v_i I(\delta-1, 2k v_i-2)\Bigr) = 0
}
\ean
as expected from the supersymmetry argument in the previous paragraph. 

The amplitude with the triplet states and two
matter fields is given by 
\ban
{\cal A}(\vec{\rho}_k, \phi_1, \phi_2) &=& 
\delta \Bigl( \rho^3_k (\xi^1 {\xi^{\bar{1}}}+\xi^2 {\xi^{\bar{2}}}) +
\rho^+_k \xi^1
\xi^2 + \rho^-_k \xi^{\bar{1}} {\xi^{\bar{2}}} \Bigr)\nonumber  \\[-2mm]
&& \; \times \; {\rm tr}({\g}_9^{k}{\lambda}_1^\dag {\lambda}_2)
\, I(\delta, 2k v_i-1)
\ean
up to a numerical
overall normalization which depends on $k$, and comes from the
contractions of the twist fields fixed at the points $i$ and $-i$ on
the double cover of the disk. Using the explicit expression of
$I(\delta, 2k v_i-1)$ and 
expanding the $\Gamma$ functions in $\delta$, 
we obtain a tree-level FI
coupling between the twisted triplet and bilinears 
in the charged matter fields. Moreover, this amplitude 
also predicts the existence of a kinetic term coupling, and an
infinite tower of derivative corrections as usual in string
theory. However, these terms disappear when we take the on-shell
limit ($\delta \rightarrow 0$) of the amplitude, so it is not safe to
extrapolate to these orders. One the other hand, this procedure can be
justified as in 
\cite{DIS} for the momentum-independent term.

A final remark to conclude this section~: the same method should allow
us to recover the tree-level couplings between twisted field and gauge
fields which were 
obtained in \cite{ABD} by factorizing the one-loop amplitude in a
background field; such direct tree-level calculation also clarifies
the fact that, in the NS-NS sector, only the singlet propagates 
between branes in the magnetic field, a result which was not
obvious within the factorization approach. 
However, as said before, one cannot really trust this computation 
since the amplitude vanishes on-shell. An alternative and more
justifiable way to obtain this coupling is to use the background field
method again. Let us just outline the procedure. The idea is to use a
boundary state which corresponds to a brane in a constant magnetic 
field on the orbifold. Those can be 
constructed directly to reproduce the amplitudes given in \cite{ABD}
or, in the alternative T-dual picture, they can be obtained by a
rotation in the spacetime directions of the twisted boundary states of
\cite{DG,T}. Then, the coupling of the twisted moduli to the magnetic
field are calculated by evaluating the scalar product of these moduli
with the boundary state. This argument
shows that in fact, in the NS-NS sector, only
the singlet couples to the magnetic field at quadratic order.

\section{ ${\cal N}=1$~: anomalous dimensions and threshold corrections}

\subsection{Anomalous dimensions~: the ${\mathbb Z}_3$ model}

In this section, we will derive the kinetic terms at one-loop for the
charged matter fields in a ${\mathbb Z}_N$ orbifold with $N$ a prime integer. Since 
there are no order two twist elements, these compactifications have only
${\cal N}=1$ sectors and no D5-branes. 
The one-loop two-point function is given by the sum of the annulus 
and M\"obius strip amplitudes:
\ban
{\cal A}(\phi^i,{\phi^{\bar{\imath}}})  
\equiv -\frac{1}{2N}\;\sum_{k=1}^{N-1}
\int_0^{i \infty} \;\frac{d\tau}{\tau} 
\left({\cal A}^{(k)}_{99}(q) + {\cal M}^{(k)}_{9}(-q)\right)
\ean
We have omitted the $k=0$ sector, which corresponds to the contribution of 
the ${\cal N}=4$ supersymmetric open string spectrum and therefore
does not contribute to wave function renormalization.
Using the ``amplitude toolbox'' given in the appendix, the two-point
function becomes 
\ba
\eqalign{
{\cal A}^{(k)}_{99}(q) \!=\! -
\frac{\delta \; \xi^i {\xi^{\bar{\imath}}}}{4\pi^2}\;
& {\rm tr}({\g}_9^{k}{\lambda}_1^\dag {\lambda}_2)\;
{\rm tr} ({\g}_9^{k})\; 
\prod_{j=1}^{3} (-2\sin {\pi k v_j}) \cr
& \times \frac{1}{2\tau}\;
\frac{\eta(\tau)^{3(1+\delta)}}
     {\vartheta[{1/2 \atop {1/2 + k v_i}}](0|\tau)} 
\int_{0}^{\tau} d\nu\; e^{i\pi \delta \frac{\nu^2}{\tau}} \;
\frac{\vartheta[{1/2 \atop {1/2 + k v_i}}](\nu|\tau)}
     {\vartheta_1(\nu|\tau)^{\delta+1}}
}\label{annulusnodd}
\ea
for the annulus contribution and
\ba
\eqalign{
{\cal M}^{(k)}_{9}(-q) \!=\!  
\frac{\delta \; \xi^i {\xi^{\bar{\imath}}}}{4\pi^2}\;
& {\rm tr}({\g}_9^{2k}{\lambda}_1^\dag {\lambda}_2)\; 
\prod_{j=1}^{3} (-2\sin {\pi k v_j}) \cr
& \times \frac{1}{2\tau}\;
\frac{\eta(\tau+1/2)^{3(1+\delta)}}
     {\vartheta[{1/2 \atop {1/2 + k v_i}}](0|\tau+1/2)} 
\int_{0}^{2\tau} d\nu\; e^{i\pi \delta \frac{\nu^2}{\tau}} \; 
\frac{\vartheta[{1/2 \atop {1/2 + k v_i}}](\nu|\tau+1/2)}
     {\vartheta_1(\nu|\tau+1/2)^{\delta+1}}
} \label{moebiusnodd}
\ea
for the M\"obius strip amplitude. We observe that the string
oscillators do not decouple and, therefore, 
contribute to the renormalization. 

Now, we will compare this result with the field theory prediction of
the anomalous dimensions for the ${\mathbb Z}_3$ model \cite{ABPSS}.  This
orientifold is defined by the twist vector ${\mathbf v} = (1/3, 1/3,
-2/3)$. The comparison requires extracting the infrared contribution
of the 
string amplitudes (\ref{annulusnodd}) and (\ref{moebiusnodd}) in the
open channel. To do this, we write the theta functions as products and
take the limit $q=e^{2 i \pi \tau} \rightarrow 0$. Expanding the
integrand in series of powers of $\delta$ (recall that we are only
interested in order one in $\delta$), the integral over $\nu$ can 
be done explicitly. Define $\alpha_i = e^{2\pi i kv_i}$. The result for
the annulus is
\ban
\lim_{q \rightarrow 0}{\cal A}^{(k)}_{99}(q) &=&  
- \frac{\delta \; \xi^i {\xi^{\bar{\imath}}}}{8\pi^2} 
{\rm tr}({\g}_9^{k}{\lambda}_1^\dag {\lambda}_2)\; 
{\rm tr}({\g}_9^{k})
\prod_{j=1}^{3} (-2\sin {\pi k v_j}) \nonumber \\
&& \; \times \; \left[
\frac{\bar{\alpha}_i^k - 1}{2 \pi \tau \delta} 
+ \frac{i \bar{\alpha}_i^k}{1-\bar{\alpha}_i^k} 
+ {\cal O}(\delta)
\right] 
\ean
and for the M\"obius strip,
\ban
\lim_{q \rightarrow 0}{\cal M}^{(k)}_{9}(-q) &= &  
\;\frac{\delta \; \xi^i {\xi^{\bar{\imath}}}}{8\pi^2}
{\rm tr}({\g}_9^{2k}{\lambda}_1^\dag {\lambda}_2)\; 
\prod_{j=1}^{3} (-2\sin {\pi k v_j})  \nonumber \\
&& \; \times \; 
\left[ \frac{\bar{\alpha}_i^{2k}-1}{2 \pi \tau \delta} 
+ \frac{i(\bar{\alpha}_i^k+\bar{\alpha}_i^{2k})}{1-\bar{\alpha}_i^k}  
+ {\cal O}(\delta)
\right] \; .
\ean
To add these contributions, one has to rescale the modular
parameter of the M\"obius strip relative to the cylinder. The
correct rescaling is obtained by normalizing the proper time in the closed
string channel ($\ell$) through the closed string propagators
\cite{MS,BF}. The relation between $\ell$ and the proper times in the
direct channel for the annulus and M\"obius strip is
$\tau_M=\tau_A/4=1/4\ell$.   
Moreover, for the ${\mathbb Z}_3$ orientifold, ${\rm tr}({\g}_9^{k})=-4$ and
$\prod_{j=1}^{3} (-2\sin {\pi k v_j})=-(-1)^k 3\sqrt{3}$. 

With this in mind, one sees
upon adding the two contributions 
that the leading ($\delta$-independent) term vanishes. This was
already observed in \cite{P} where the one-loop FI
term was calculated in the GS formalism, and shown to vanish.
Cutting off the integral
by introducing the infrared regulator $t=-2i\tau_A \leq 1/\mu^2$, we
find the infrared behavior:  
\ba
\eqalign{
{\cal A} \! = \!  
-\frac{3i\sqrt{3}}{32\pi^2} \; 
\delta \;\xi^i {\xi^{\bar{\imath}}}\;
\sum_{k=1}^{2} \frac{1}{1-\bar{\alpha}_i^k}
{\rm tr}({\g}_9^{k}{\lambda}_1^\dag {\lambda}_2)\; 
{\rm ln}\frac{\mu^2}{M^2_I}} \; .
\label{sum}
\ea 
To evaluate the Chan-Paton trace, we need to introduce a little
bit more detail. As explained in \cite{ABPSS},
the massless matter content of the ${\mathbb Z}_3$ model is given by
three copies of the $(1, \overline{66})_{-2}$ and
$(8, 12)_1$ representations of an $SO(8) \times SU(12) \times U(1)$ 
gauge group (the superscripts denote the $U(1)$ charge). 

To be explicit, we introduce the generators $\sigma_{ar}$ and 
$\tau_{rs}$, for $a=1,\cdots,8$ and $r,s=1,\cdots,12$, normalized such
that ${\rm tr}({\sigma^T_{ar} 
\sigma_{bs}})= \frac{1}{2}\, \delta_{ab} \delta_{rs}$ 
and ${\rm tr}({\tau_{pq} \tau_{rs}})= \frac{1}{2}\, (\delta_{ps}
\delta_{qr} - 
\delta_{pr} \delta_{qs})$. In this basis, the matter fields can be
written as $\phi^i \lambda_{i} = 2 \psi^{ir}_a \sigma_{ar}$
for the $(8, 12)$ representation and $\phi^i \lambda_{i} =
2 \chi^i_{rs}\tau_{rs}$ for the $(1,\overline{ 66})$. 
Starting from the ten dimensional SYM theory and 
performing the Kaluza-Klein reduction to four dimensions, this
gives correctly normalized kinetic terms for the complex
matter fields. 
The Chan-Paton trace in (\ref{sum}) can now be evaluated:
\ba
\Delta {\cal L}^{\rm (S)}_{\rm one-loop} = 
\frac{3}{32 \pi^2}\;{\rm ln}\frac{\mu^2}{M^2_I}\; 
   (\partial_\mu \psi^{ir}_a \partial^\mu {\psi^{ir\dag}_a} 
- 2 \partial_\mu \chi^i_{rs} \partial^\mu {\chi^{i \dag}_{rs}})  
\label{string}
\ea
for the one-loop renormalization of these fields in the string
frame. To compare this result with the field theory prediction, we 
first go to the Einstein frame. The IR divergent terms can therefore
be summarized by the Lagrangian (\ref{oneloop}), with 
$\tilde\gamma_{\psi} =\frac{3}{32\pi^2}\, {\rm ln}({\mu^2}/{M^2_I})$ and
$\tilde\gamma_{\chi} =-\frac{3}{16\pi^2}\, {\rm
ln}({\mu^2}/{M^2_I})$. Now, we will 
show that these coefficients are related to the anomalous dimensions
$\gamma$ of the matter fields according to 
$\gamma\; {\rm ln}({\mu^2}/{M^2_I})=\tilde\gamma/{\rm Im}\, S$.

In an ${\cal N}=1$ SYM theory with a simple gauge group and a generic 
superpotential,
$${\cal W}= \frac{1}{6} 
\lambda_{ijk}^{abc} \phi_a^i \phi_b^j \phi_c^k,
$$
where $i,j,k$ are family indices and $a, b, c$ group indices 
(for us, the family indices will label the three complex planes), 
the anomalous dimensions of the matter fields $\phi^i_a$ are given by
the formula \cite{W}  
$${(\gamma_i^{\ j})}^a_{\ b} = -\frac{1}{16\pi^2}(2 g^2 C_2(R_a)\;
\delta_i^{\ j} \delta^a_{\ b} - \sum_{kl,\; cd} \lambda_{ikl}^{acd}
\lambda^{jkl}_{bcd})$$ 
where $C_2(R_a)$ is the quadratic Casimir of the representation $R_a$
and $\lambda^{ijk}_{abc} = {\lambda_{ijk}^{abc}}^*$.
This formula can be easily generalized to semi-simple groups with
$U(1)$ factors. In particular, for the ${\mathbb Z}_3$ model,
where the superpotential is given by
$${\cal W}=\sqrt{\frac{1}{2\,{\rm Im}\, S}}\; \epsilon_{ijk}\;
\psi^i \chi^j \psi^k$$ 
one finds
\ban
(\gamma_\psi)_i^{\ j} &=& -\frac{1}{16\pi^2}\left({(2 g^2_{SU(12)}
C^{SU(12)}_2(12) + 2 g^2_{SO(8)} C^{SO(8)}_2(8) + 
g^2_{U(1)}) - \frac{11}{{\rm Im}\, S} }\right) \delta_i^{\ j}   \\
(\gamma_\chi)_i^{\ j} &=& -\frac{1}{16\pi^2}\left({(2 g^2_{SU(12)}
C^{SU(12)}_2(66) + 4 g^2_{U(1)}) - \frac{8}{{\rm Im}\, S} }\right)
\delta_i^{\ j} 
\ean
where we have suppressed a multitude of
Kronecker deltas in the group indices.
The coupling constants for the gauge groups are
$g^2_{SO(8)}=1/{\rm Im}\, S, \; g^2_{SU(12)}=1/(2\,{\rm Im} \, S), \;
g^2_{U(1)}=1/(24\,{\rm Im} \, S)$ so the final result is
\be
{(\gamma_{\psi})}_i^{\ j} =\frac{3 \, \delta_i^{\ j}}{32\pi^2\,
{\rm Im} \, S},
   \qquad
{(\gamma_{\chi})}_i^{\ j} = -\frac{3 \, \delta_i^{\ j}}{16\pi^2\,
{\rm Im} \, S} ,
\ee
in agreement with the string theory result.

\subsection{One-loop Fayet-Iliopoulos term~: the ${\mathbb Z}_6'$ model}

The ${\mathbb Z}_6'$ model \cite{AFIV} is defined by the twist vector ${\mathbf 
v}=(1/6, -1/2, 1/3)$. 
The requirement of tadpole cancellation forces us to
introduce 32 D9-branes and also 32 D5-branes filling the space transverse 
to the first and second complex planes. We refer the reader to section 
five of \cite{ABD} for more details on this model;  here we only
summarize the 
characteristics needed to compute the one-loop kinetic 
term of the matter fields. The model contains an ${\cal N}=4$
sector $(\theta^0)$, two  
${\cal N}=1$ sectors $(\theta^1,\theta^5)$ and three ${\cal N}=2$ sectors
$(\theta^2,\theta^3,\theta^4)$. 
For $k=2,4$ (respectively $k=3$), the 
second (resp. third) complex plane is untwisted.
It is important to note that since the D5-branes fill the third
complex plane but not the first and the second,
$9-5$ strings in the $k=3$ sector enjoy the full ${\cal N}=2$
supersymmetry, whereas $9-5$ strings in the $k=2,4$ sectors 
see ${\cal N}=1$ supersymmetry only. In the former sector,
the D5-branes are transverse only to twisted directions, and thus
break no supersymmetry that was not already broken by 
the D9-branes and the action of the orientifold.

The amplitudes for the two-point function of the 
scalar $\phi^1$ are
\ban
\eqalign{
{\cal A}_{99}(\phi^1,{\phi^{\bar 1}}) \!=\! 
\frac{\delta \;\xi^1 \xi^{{\bar 1}}}{96\pi^2} 
&\sum_{k=1,5} 
{\rm tr}({\g}_9^{k}{\lambda}_1^\dag {\lambda}_2)\;
{\rm tr}({\g}_9^{k})\; 
\prod_{j=1}^{3} (-2\sin {\pi k v_j}) \cr  
& \times \int_0^{i \infty} \;\frac{d\tau}{\tau^2}\;
\frac{\eta(\tau)^{3(1+\delta)}}
     {\vartheta[{1/2 \atop {1/2 + k/6}}](0|\tau)} \;
\int_{0}^{\tau} d\nu\; e^{i\pi \delta \frac{\nu^2}{\tau}} \; 
\frac{\vartheta[{1/2 \atop {1/2 + k/6}}](\nu|\tau)}
     {\vartheta_1(\nu|\tau)^{\delta+1}}, \cr
{\cal M}_{9}(\phi^1,{\phi^{\bar 1}}) \!=\! \; - 
\frac{\delta \;\xi^1 \xi^{{\bar 1}}}{96\pi^2} 
&\sum_{k=1,5} 
{\rm tr}({\g}_9^{2k}{\lambda}_1^\dag {\lambda}_2)\; 
\prod_{j=1}^{3} (-2\sin {\pi k v_j}) \cr  
& \times \int_0^{i \infty} \;\frac{d\tau}{\tau^2}\;
\frac{\eta(\tau+1/2)^{3(1+\delta)}}
     {\vartheta[{1/2 \atop {1/2 + k/6}}](0|\tau+1/2)} \;
\int_{0}^{2\tau} d\nu\; e^{i\pi \delta \frac{\nu^2}{\tau}} \; 
\frac{\vartheta[{1/2 \atop {1/2 + k/6}}](\nu|\tau+1/2)}
     {\vartheta_1(\nu|\tau+1/2)^{\delta+1}}, \cr
{\cal A}_{95}(\phi^1,{\phi^{\bar 1}}) \!=\! - 
\frac{\delta \;\xi^1 \xi^{{\bar 1}}}{48\pi^2} 
&\sum_{k=1,2,4,5} 
{\rm tr}({\g}_9^{k}{\lambda}_1^\dag {\lambda}_2)\;
{\rm tr}({\g}_5^{k})
\sin \frac{\pi k}{3} \cr  
& \times \int_0^{i \infty} \;\frac{d\tau}{\tau^2}\;
\frac{\eta(\tau)^{3(1+\delta)}}
     {\vartheta[{0 \atop {1/2 + k/6}}](0|\tau)} \;
\int_{0}^{\tau} d\nu\; e^{i\pi \delta \frac{\nu^2}{\tau}} \; 
\frac{\vartheta[{0 \atop {1/2 + k/6}}](\nu|\tau)}
     {\vartheta_1(\nu|\tau)^{\delta+1}}.
}
\ean
This field $\phi^1$, which comes from a
complex plane twisted by all sectors of
the orbifold (ie.\ $kv_1 \notin {\mathbb Z}$ for all $k$), 
only receives contributions from the ${\cal N}=1$
sector for the ${\cal A}_{99}$ and ${\cal M}_9$ amplitudes. For these two
diagrams, the contributions of the ${\cal N}=2$ sectors
vanish, just like we already observed for the scalars of the ${\cal
N}=2$ hypermultiplets in $K3 \times T^2$ compactifications. 
Notice that, due to the tadpole conditions ${\rm tr}({\g}_9^{k})=
{\rm tr}({\g}_5^{k})=0$ for $k=1, 3, 5$, the amplitude ${\cal 
A}_{99}$ vanishes identically, and one immediately sees that
there is no one-loop FI terms 
proportional to ${\rm tr}({\g}_9^{k}{\lambda}_1^\dag {\lambda}_2)$
with $k$ odd, since the M\"obius strip can only contribute to even 
powers in $\gamma^k$. Further, one can perform an expansion as already
performed for the ${\mathbb Z}_3$ model, using now the conditions ${\rm
tr}({\g}_5^{2})=-{\rm tr}({\g}_5^{4})=-8$ and ${\g}_9^{6}=-1$
\cite{AFIV}. The result is
that the contribution of the ${\cal N}=1$, $k=1,5$
sectors of the M\"obius strip to the would-be one-loop FI term is
cancelled by the $k=2,4$ sectors of the $9-5$ annulus, which 
are actually ${\cal N}=1$ for this $9-5$ amplitude only, as noted
above.

The FI D-term would have looked like $\zeta_{\rm FI}^2 \sim
\Lambda_{\rm UV} \, {\rm tr} \, Q_{U(1)}$ for a UV cutoff
$\Lambda_{\rm UV}$, which would have generated a mass term for
the charged scalar, proportional to its $U(1)$ charge.
We thus see that this term is not generated.
One can easily check that such mass terms are also absent
for the two other scalar fields, $\phi^2$ and $\phi^3$. 
Finally, as in the previous section, the one-loop renormalization of this
field is given by its field theoretical ${\cal N}=1$ anomalous
dimensions.

\subsection{Threshold corrections~: the ${\mathbb Z}_6'$ model}

The scalar $\phi^2$ and its complex conjugate $\phi^{\bar 2}$ come 
from a plane which is untwisted in the $k=2,4$ sectors. The relevant 
one-loop two-point amplitudes are
\ban
\eqalign{
{\cal A}_{99}(\phi^2,{\phi^{\bar 2}}) \!=\! 
\frac{\delta \;\xi^2 \xi^{{\bar 2}}}{96\pi^2} &\left[  
\sum_{k=1,5} 
{\rm tr}({\g}_9^{k}{\lambda}_1^\dag {\lambda}_2)\;
{\rm tr}({\g}_9^{k})\; 
\prod_{j=1}^{3} (-2\sin {\pi k v_j})  \right. \cr 
& \times  \int_0^{i \infty} \;\frac{d\tau}{\tau^2} \; 
\frac{\eta(\tau)^{3(1+\delta)}}
     {\vartheta[{1/2 \atop {1/2 - k/2}}](0|\tau)} \; 
\int_{0}^{\tau} d\nu\; e^{i\pi \delta \frac{\nu^2}{\tau}} \;
\frac{\vartheta[{1/2 \atop {1/2 - k/2}}](\nu|\tau)}
     {\vartheta_1(\nu|\tau)^{\delta+1}} \cr
+  & \left. \sum_{k=2,4} 
{\rm tr}({\g}_9^{k}{\lambda}_1^\dag {\lambda}_2)\;
{\rm tr}({\g}_9^{k})\; 
\prod_{j=1,3} (2\sin {\pi k v_j}) 
\int_0^{i \infty} \;\frac{d\tau}{\tau}\;\Gamma_2^{(2)}(\tau) \right], \cr
{\cal M}_{9}(\phi^2,{\phi^{\bar 2}}) \!=\! - 
\frac{\delta \;\xi^2 \xi^{{\bar 2}}}{96\pi^2}  & \left[ 
\sum_{k=1,5} 
{\rm tr}({\g}_9^{2k}{\lambda}_1^\dag {\lambda}_2)\; 
\prod_{j=1}^{3} (-2\sin {\pi k v_j}) \right. \cr  
& \times  \int_0^{i \infty} \;\frac{d\tau}{\tau^2}\; 
\frac{\eta(\tau+1/2)^{3(1+\delta)}}
     {\vartheta[{1/2 \atop {1/2 - k/2}}](0|\tau+1/2)} \;
\int_{0}^{2\tau} d\nu\; e^{i\pi \delta \frac{\nu^2}{\tau}} \;
\frac{\vartheta[{1/2 \atop {1/2 - k/2}}](\nu|\tau+1/2)}
     {\vartheta_1(\nu|\tau+1/2)^{\delta+1}} \cr 
+  \; 2 & \left. \; \sum_{k=2,4} 
{\rm tr}({\g}_9^{2k}{\lambda}_1^\dag {\lambda}_2)\; 
\prod_{j=1,3} (2\sin {\pi k v_j}) \int_0^{i \infty}\;
\frac{d\tau}{\tau}\;\Gamma_2^{(2)}(\tau) \right], \cr
{\cal A}_{95}(\phi^2,{\phi^{\bar 2}}) \!=\! - 
\frac{\delta \;\xi^2 \xi^{{\bar 2}}}{48\pi^2} & 
\sum_{k=1,2,4,5} 
{\rm tr}({\g}_9^{k}{\lambda}_1^\dag {\lambda}_2)\;
{\rm tr}({\g}_5^{k})
\sin \frac{\pi k}{3} \cr  
& \times \int_0^{i \infty} \;\frac{d\tau}{\tau^2}\;
\frac{\eta(\tau)^{3(1+\delta)}}
     {\vartheta[{0 \atop {1/2 - k/2}}](0|\tau)} \;
\int_{0}^{\tau} d\nu\; e^{i\pi \delta \frac{\nu^2}{\tau}} \;
\frac{\vartheta[{0 \atop {1/2 - k/2}}](\nu|\tau)}
     {\vartheta_1(\nu|\tau)^{\delta+1}}.
}
\ean
We observe that, besides the field theoretical ${\cal N}=1$ 
renormalization running up to the string scale, the corrections given 
by the ${\cal N}=2$ sectors depend on the geometric moduli of the
complex planes in the same way as for the gauge bosons. Indeed, in the 
sectors $k=2,4$
where the scalars are untwisted, one can use the background
field method suggested at the beginning of section \ref{sec:general}
to obtain a result
identical to that of the gauge bosons. Notice that this argument
also shows that at tree-level,
the twisted NS-NS field in the $k=2,4$ sectors
couples to the kinetic term of this matter field $\phi^2$ 
in the same way
as the gauge field does. We can explain 
this phenomenon as follows: we start with a four-dimensional 
orientifold compactification with a twist vector defined as 
${\mathbf v^\prime}=2{\mathbf v}=(1/3, -1, 2/3)$ which generates a 
${\mathbb Z}_{3}$ subgroup of the original ${\mathbb Z}^\prime_{6}$. The result is 
an ${\cal N}=2$, ${\mathbb Z}_3$ orbifold of the family we studied in the 
first part of this section. However, it will also have D5-branes
which, as we have already seen, are crucial for the absence of
one-loop FI terms, but do not fill completely the space transverse to
$K3$.
This ${\cal N}=2$ orbifold leaves the second complex plane untwisted. 
As argued above, the scalar fields associated to this plane belong 
to vector multiplets, and so they have the same one-loop 
renormalization and couplings to the twisted fields of the orbifold. 
The projection on ${\mathbb Z}_6^\prime$ invariant states eliminates some 
of the fields but these couplings survive. 

We now describe in more detail the threshold corrections coming from 
${\cal N}=2$ sectors. We denote by $U$ the complex
structure of the second two-torus:
$$U=\frac{G_2^{12}+i\sqrt{G_2}}{G_2^{11}}.$$
The threshold corrections due to 
${\cal N}=2$ sectors are given by the sum of ${\cal A}^{(k)}_{99}$ 
and ${\cal M}^{(k)}_{9}$ for $k=2,4$:
\ba\eqalign{
\label{thresholds}
\frac{1}{6(2\pi)^2}\sum_{k=2,4}& \prod_{j=1,3} (\sin {\pi k v_j}) 
\left[{\rm tr}({\g}_9^{k}{\lambda}_1^\dag {\lambda}_2)\,
{\rm tr}({\g}_9^{k}) - 
2\, {\rm tr}({\g}_9^{2k}{\lambda}_1^\dag {\lambda}_2) \right] \int
\frac{d\tau}{\tau} \Gamma_2^{(2)}(\tau) \cr
&= \frac{3}{(4\pi)^2}\sum_{k=2,4}
{\rm tr}({\g}_9^{k}{\lambda}_1^\dag {\lambda}_2)
\left[\ln\left(\sqrt{G_2}\, {\rm Im} \, U \, \mu^2\right) + 
4\;{\rm Re} \ln \eta(U)
\right] \; .
}
\ea
If the Chan-Paton matrices $\lambda_i$ had been in the adjoint
representation of 
the group, the coefficients of these threshold corrections would have
been the ${\cal N}=2$ effective theory beta functions of the
corresponding sector. These corrections reproduce the
heterotic ones only in the limit ${\rm Im}\, T \rightarrow
0$. Non-perturbative corrections are needed to reproduce the
complete threshold dependence on $T$ \cite{AGNT} which, on
the heterotic side, is just the K\"ahler modulus of the torus, but on
the type I side depends on the ten-dimensional string coupling
constant as $T=b_2^{{\rm R-R}}+ i \sqrt{G_2} M_{\rm S}^2 e^{-\Phi_{10}}$.

We have obtained similar results for the scalar $\phi^3$,
except for one important point, its tree-level couplings to the
twisted moduli of the ${\cal N}=2$ sector.  Indeed, as noticed in
\cite{ABD}, the corresponding twisted field belongs  
to a hypermultiplet --- which cannot couple to kinetic terms of
 non-abelian gauge fields because of ${\cal N}=2$ supersymmetry ---
and the complex field $\phi^3$ is in the vector multiplet of the 
${\cal N}=2$, ${\mathbb Z}_2$ orientifold generated by ${\mathbf v^\prime}=
3{\mathbf v}=(1/2,-3/2,1)$. On the other hand, it can couple to ${\cal
N}=1$ twisted fields; such couplings should be obtained using the method
described in the final part of the previous section. 
Finally, using tadpole conditions, one shows that its threshold
corrections come from the ${\cal M}^{(3)}_{9}$ 
and  ${\cal A}^{(0)}_{95}$ amplitudes and depends on the complex
structure of the third two-torus, a result
which can be explained as for $\phi^2$.

\section{\bf Conclusions and discussion}

In this article, we have investigated some parts of the effective action
of four-dimensional type I compactifications, focusing in particular
on the one-loop renormalization of the K\"ahler metric and the
tree-level couplings between charged matter fields and twisted
moduli. 

For the renormalization of the K\"ahler metric, the general picture is
the following: on the one hand, ${\cal N}=1$ sectors yield moduli-independent
corrections to the metric, and hence to the physical Yukawa
couplings. Due to the reduced number of supersymmetries, the string
oscillators do not decouple, and the renormalization
constant of the charged field is given by infrared logarithmic
corrections, independent of the volume of the compact space,
cut off at the string scale $M_{\rm S}$ and with a coefficient given
by the field theory $\gamma$ functions.
On the other hand, moduli-dependent threshold corrections
arise in ${\cal N}=2$ sectors, for scalar fields associated to the
plane left invariant by the twist operator in these sectors. 
The phenomenological
use of this kind of corrections in models with low string scale 
is discussed in \cite{B}.
For  a rectangular untwisted torus of radii $R_1$, $R_2$,
these corrections are proportional to 
${\rm ln}\, (\mu^2 R_1R_2) + f(R_1/R_2)$
where $\mu$ is the infrared scale and 
$f$ diverges linearly when $R_1 >> R_2$.
Otherwise, the scalars associated to twisted plane are not
renormalized. The contribution of the D9-D5 annulus amplitude is
special, in the sense that in the ${\cal N}=2$ sectors where the
D5-brane is wrapped around twisted directions and therefore breaks
half of these ${\cal N}=2$ supersymmetries, it gives corrections
similar to those of ${\cal N}=1$ sectors.

Within this computation of one-loop amplitudes, we have also recovered
the result of \cite{P} on the absence of one-loop induced
Fayet-Iliopoulos 
term and generalized it to ${\cal N}=1$ orientifolds with ${\cal N}=2$
sectors and D5-branes. This vanishing occurs because of the
cancellation between 
contributions of worldsheets with different topology and of different
sectors. In particular, the presence of D5-branes is crucial in this
mechanism. This cancellation is related to the absence of twisted R-R
tadpoles in these models. 

Finally, we have calculated explicitly tree-level couplings between
the twisted fields of the orbifold and charged fields for
$K3 \times T^2$ orientifolds. In agreement to supersymmetry
predictions, we have obtained couplings between charged
matter fields and the three 
NS-NS twisted moduli which, with a R-R twisted scalar, make up a
hypermultiplet and which transform as a
triplet of the $R$-symmetry group $SU(2)_R$. These NS-NS fields are
also the blow-up modes of the orbifold.
On the other hand, the coupling constant contains 
a tree-level part proportional to the scalar component of ${\cal
N}=2$ twisted vector-tensor multiplet, which is also a singlet under 
$SU(2)_R$.
The CP-odd counterpart of these couplings have been
investigated in detail in \cite{SS1}, where they were extracted 
by factorization of one-loop amplitudes in the odd spin-structure. The
result is that the ${\rm tr}(F\wedge F)$ couples to the twisted R-R 
tensor which is in a $D=6,\, {\cal N}=(1,0)$ tensor multiplet while
the $U(1)$ field couples to the R-R scalar field as $\int 
{^6C_k^{(0)}}{\rm tr}\gamma^{(k)} F$. The supersymmetric partner of the
Chern-Simons coupling of the tensor field is the
tree-level coupling between the singlet $b_k^{(0)}$ and $F^2$ while the 
counterpart of the other term is 
given by the Fayet-Iliopoulos D-terms~: $\int \vec{\rho} \cdot \vec{D}$. 
Integrating out the auxiliary fields $\vec{D}$ gives, as explained in
\cite{DM}, the coupling between bilinears in the 
charged matter field and the twisted triplet that we have calculated
directly in string theory.  

Such disk calculations should easily generalize to ${\cal N}=1$
compactifications, for which the twisted moduli space structure was
described in \cite{Kl}, for instance. Again, the CP-odd partners of
these couplings have been analyzed in \cite{SS2}; for twisted sectors
without fixed plane, the closed string twisted fields belong to linear
multiplets, and its R-R part couples as Green-Schwarz terms to $U(1)$
gauge fields and $F\wedge F$. By supersymmetry, we also expect FI
couplings for its NS-NS partner.
The ${\cal N}=2$ sectors are more involved. Actually, the ambiguity
raised in \cite{SS2}, where 
they were unable to fix completely the anomalous couplings through the
factorization approach, should be determined by the disk
calculation. As a final comment, we evoke an open issue in these
orientifold compactifications: the problem of target space duality
symmetry \cite{IRU,Kl,LLN} is not completely settled and needs more
investigation.

\acknowledgments

We would like to thank C. Bachas for collaboration on this project and
helpful discussions and suggestions, and C. Angelantonj
for useful discussions. PB thanks E. Kiritsis for a discussion.
PB is financially supported by the ``Minist\`ere de l'Equipement,
des Transports et de l'Am\'enagement du Territoire''.
MB is financially supported by the Swedish Institute and the
Royal Swedish Academy of Sciences.

\vfill
\eject
\setcounter{section}{0}   

\Appendix{Amplitude toolbox}
\label{app:toolbox}

\centerline{\bf One-loop partition functions}

The contribution of the zero modes and oscillators of the 
spacetime coordinates and ghosts to the partition function, common to 
all the compactifications considered in this paper, is 
\ba
Z_{4}^{\a,\b}(\tau) \!=\! \frac{1}{4\pi^4\tau^2} 
\frac{\vartheta[{\a \atop \b}](0|\tau) }{\eta(\tau)^3}
\ea
for the annulus. 

\noindent For the $K3 \times T^2$ orientifolds, the internal annulus
partition function is~: 
\ba
Z_{{\rm int},\; k}^{\a,\b}(\tau) \!=\! -\; \Gamma^{(2)}(\tau) \;
\frac{\vartheta[{\a \atop \b}](0|\tau) }{\eta(\tau)^3}  \;
(2\sin {\pi k\over N})^2 \; \prod_{j=1,2} 
\frac{\vartheta[{\a \atop {\b + kv_j}}](0|\tau)}
{\vartheta[{1/2 \atop {1/2 +kv_j}}](0|\tau) }
\ea
where $\Gamma^{(2)}(\tau)$ is the lattice sum over momenta along the
untwisted two-torus:
\ba
\Gamma^{(2)}(\tau) = \sum_{n_4,n_5} e^{2i \pi \tau |n_4+n_5 U|^2 /
(\sqrt{G}\, {\rm Im} \, U)}
\ea
with $G_{ab} \, (a,b=4,5)$ the torus metric, and
$U=(G_{45} + i\sqrt{G})/G_{44}$ its complex structure. 

\noindent For the $T^6/{\mathbb Z}_3$ model, the internal annulus partition
function is~:
\ba
Z_{\rm int,\; k}^{\a,\b}(\tau) \!=\! \prod_{j=1}^{3} 
(-2\sin {\pi k v_j}) \; 
\frac{\vartheta[{\a \atop {\b + k v_j}}](0|\tau)}
     {\vartheta[{1/2 \atop {1/2 + k v_j}}](0|\tau) }
\ea
For the $T^6/{\mathbb Z}_6^\prime$ model, the internal annulus partition
functions are~:
\ba
\eqalign{
Z_{{\rm int},\; k}^{\a,\b}(\tau) \!&=\! 
\prod_{j=1}^{3} (-2\sin {\pi k v_j}) \;
\frac{\vartheta[{\a \atop {\b + k v_j}}](0|\tau)}
     {\vartheta[{1/2 \atop {1/2 + k v_j}}](0|\tau) }, \ \ k=1,5 \cr  
Z_{{\rm int},\; k}^{\a,\b}(\tau) \!&=\! \Gamma_2^{(2)}(\tau) \;
\frac{\vartheta[{\a \atop {\b + kv_2}}](0|\tau)}
     {\eta(\tau)^3}  \; 
\prod_{j=1,3} (2\sin {\pi k v_j}) \; 
\frac{\vartheta[{\a \atop {\b + k v_j}}](0|\tau)}
     {\vartheta[{1/2 \atop {1/2 + k v_j}}](0|\tau) }, \ \ k=2,4 \cr
Z_{{\rm int},\; k}^{\a,\b}(\tau) \!&=\! \Gamma_3^{(2)}(\tau) \;
\frac{\vartheta[{\a \atop {\b + 3v_3}}](0|\tau)}
     {\eta(\tau)^3}  \;
\prod_{j=1,2} (2\sin {3\pi v_j}) \; 
\frac{\vartheta[{\a \atop {\b + 3v_j}}](0|\tau)}
     {\vartheta[{1/2 \atop {1/2 + 3v_j}}](0|\tau) }, \ \ k=3 \cr
}
\ea
The internal partition functions for the M\"obius strip and the 
annulus with one boundary on a D9-brane and the other on a D5-brane 
can be found in appendix 2 of \cite{ABD}.

\eject

\centerline{\bf One-loop correlation functions}

The bosonic correlation function on the torus ${\cal T}$
in the untwisted directions is: 
\ba
\langle X(z_1) X(z_2)\rangle _{{\cal T}} = 
-\frac{1}{4} \, \ln \, {\left| 
\frac{\vartheta_{1}(\nu_1-\nu_2|\tau)}{\vartheta^\prime_{1}(0|\tau)} 
\right|}^{2} + 
\frac{\pi\,  ({\rm Im}(\nu_1-\nu_2))^2}{2 \, {\rm Im}\,  \tau} \; ,
\ea 
The correlators on the annulus ${\cal A}$
and the M\"obius strip ${\cal M}$ are obtained 
by symmetrizing this function under the involutions
\ba
I_{{\cal A}}(\nu) = I_{{\cal M}}(\nu) = 1 - {\bar \nu} \; .
\ea
The fermionic correlation functions on the torus are
\ba
\eqalign{
\langle \psi^\mu(z_1)\psi^\nu(z_2)\rangle ^{\a,\b}_{{\cal T}} =
\frac{i}{2}\;  
\frac{\vartheta[{\a \atop \b}](\nu_1-\nu_2 |\tau) 
{\vartheta_1^\prime}(0|\tau)}
{\vartheta[{\a \atop \b}](0|\tau)  
\vartheta_1(\nu_1-\nu_2 |\tau)}\; 
\delta^{\mu \nu} \; , \cr 
\langle \psi^i(z_1)\psi^j(z_2)\rangle ^{\a,\b}_{{\cal T}} = \frac{i}{2}\; 
\frac{\vartheta[{\a \atop {\b + kv_i}}](\nu_1-\nu_2 |\tau)
{\vartheta_1^\prime}(0|\tau)}
{\vartheta[{\a \atop {\b + kv_i}}](0|\tau)  
\vartheta_1(\nu_1-\nu_2 |\tau)}\; 
\delta^{i \bar\jmath} 
}
\ea 
for untwisted and twisted worldsheet fermions in the even spin
structures. Like the boson propagators, the fermion propagators on the
other  
surfaces can be determined from these correlators by the method of 
images (see appendix of \cite{ABFPT} for more details).

The correlation function of two
twisted fermions for strings with DN boundary conditions is 
\ba
\eqalign{
\langle \psi^i(z_1)\psi^j(z_2)\rangle ^{\a,\b}_{{\cal A}_{95}} = \frac{i}{2}\; 
\frac{\vartheta[{\a + 1/2 \atop {\b + kv_i}}](\nu_1-\nu_2 |\tau)
{\vartheta_1^\prime}(0|\tau)  }
     {\vartheta[{\a + 1/2  \atop {\b + kv_i}}](0|\tau)  
\vartheta_1(\nu_1-\nu_2|\tau)}\; 
\delta^{i \bar{j}}
}
\ea 

\ 

\centerline{\bf Theta function identity}

{\small{ 
\ba
\eqalign{
\sum_{\a,\b=0,1/2 \atop {\rm even}} \frac{1}{2}\; \eta_{\a,\b}\; & 
\vartheta \left[{\a \atop {\b}}\right](\nu|\tau)\;  
\vartheta \left[{\a + \d_1 \atop {\b + \gamma_1}}\right](\nu|\tau)\;  
\vartheta \left[{\a + \d_2 \atop {\b + \gamma_2}}\right](0|\tau)\;  
\vartheta \left[{\a + \d_3 \atop {\b + \gamma_3}}\right](0|\tau) =
\cr  
& \frac{1}{2} \; 
\vartheta \left[{1/2 \atop {1/2}}\right](\nu|\tau)\;  
\vartheta \left[{1/2 + \d_1 \atop {1/2 + \gamma_1}}\right](\nu|\tau)\;    
\vartheta \left[{1/2 + \d_2 \atop {1/2 + \gamma_2}}\right](0|\tau)\;  
\vartheta \left[{1/2 + \d_3 \atop {1/2 + \gamma_3}}\right](0|\tau)
} \label{thetaid}
\ea
}}
valid for 
$\d_1 + \d_2 + \d_3 = 0$ and $\gamma_1 + \gamma_2 + \gamma_3 = 0$.

\ 

\centerline{\bf Tree-level twisted correlation functions}

The twist field correlators on the disk are~:
\ba
\eqalign{
\label{bosons}
\langle \sigma_k\psi^m(z_1)& \psi^i(x_1) {\psi^j}(x_2) 
\psi^n\sigma^\dag_k(z_2)\rangle  =
{\left({\frac{z_1-x_1}{z_2-x_1}}\right)}^{k v_i}
{\left({\frac{z_1-x_2}{z_2-x_2}}\right)}^{k v_j} \cr
&\times \Bigl(
 \frac{\delta^{m\bar{\imath}} \delta^{j\bar{n}}}{(z_1-x_1)(x_2-z_2)}
-\frac{\delta^{m\bar{\jmath}} \delta^{i\bar{n}}}{(z_1-x_2)(x_1-z_2)}
+\frac{\delta^{m\bar{n}} \delta^{i\bar{\jmath}}}{(z_1-z_2)(x_1-x_2)}
\Bigr)
}
\ea
for the worldsheet fermions and 
\ba
\eqalign{
\label{fermions}
\langle \sigma_k(z_1) \partial \phi^i(x_1) 
                      \partial \phi^j (x_2)
        \sigma^\dag_k(z_2)
\rangle  = &
{\left({\frac{z_1-x_1}{z_2-x_1}}\right)}^{-(1-k v_i)}
{\left({\frac{z_1-x_2}{z_2-x_2}}\right)}^{-k v_i} \cr
&\times \frac{\delta^{i \bar{\jmath}}}{(x_1-x_2)^2} 
{\left[{(1-k v_i)\frac{(z_1-x_1)}{(z_2-x_1)} + k v_i 
\frac{(z_1-x_2)}{(z_2-x_2)}}\right]}
}
\ea
for the bosons (up to a
$(z_1-z_2)$ dependent term which comes from the contraction of the twist
fields on the disk, 
and is the same for (\ref{bosons}, \ref{fermions})). 
We also use the method of  images to obtain the
correlation functions of both left- and right-moving fields.


\end{document}